\documentclass{PoS}
\usepackage{amsmath}
\usepackage{amsfonts}
\usepackage{amssymb}
\newcommand{\vc}[1]{{\ensuremath \vec{#1}}}

\title{The heavy quark-antiquark potential from lattice and perturbative QCD}

\ShortTitle{The heavy quark-antiquark potential from lattice and perturbative QCD}

\author{\speaker{Alexander Laschka}, Norbert Kaiser and Wolfram Weise\\
        Physik-Department, Technische Universit\"at M\"unchen, D-85747 Garching, Germany\\
        E-mail: \email{alaschka@ph.tum.de},
                \email{nkaiser@ph.tum.de},
                \email{weise@ph.tum.de}
}

\abstract{
The heavy quark-antiquark potential in perturbative QCD is subject to
ambiguities. We show how to derive a well-defined and stable short-distance
potential that can be matched to results from lattice QCD simulations at
intermediate distances. The static potential as well as the order 1/m
potential are discussed.
}

\FullConference{8th Conference Quark Confinement and the Hadron Spectrum\\
                 September 1-6, 2008\\
                 Mainz. Germany}

\begin{document}
The static quarkonium potential has been studied by lattice simulations as well
as in perturbative QCD. It is an ideal object for exploring the interplay
between perturbative and non-perturbative physics. However, the perturbative
prediction tends to fail already at very small distances. It was found that
this behaviour can be understood in the context of
renormalons~\cite{Beneke:1998rk;Hoang:1998nz}.

At two-loop order the static potential reads in
momentum space~\cite{Peter:1996ig;Peter:1997me;Schroder:1998vy}
\begin{equation*}
  \tilde{V}^{(0)}(|\vc q|) = -\frac{4\pi C_F \alpha_s(|\vc q|)}{\vc q\,^2}\,
  \bigg\{1+\frac{\alpha_s(|\vc q|)}{4\pi}\, a_1
  +\left(\frac{\alpha_s(|\vc q|)}{4 \pi}\right)^2 a_2
  + \ldots \bigg\}\, ,
\end{equation*}
where $\vc q$ is the three-momentum transfer.
Higher order terms involving IR divergences are not considered at this
point. We define the static potential in coordinate space by a restricted
Fourier transform with a low-momentum cutoff $q_{\textrm{min}}$:
\begin{equation*}
  V^{(0)}(|\vc r|) = \intop_{|\vc q|>q_{\textrm{min}}}\!\! \frac{d^3\! q}{(2\pi )^3}\ 
  e^{i\vc q\cdot\vc r}\,
  \tilde{V}^{(0)}(|\vc q|).
\end{equation*}

\begin{figure}[h]
\includegraphics[width=0.48\textwidth]{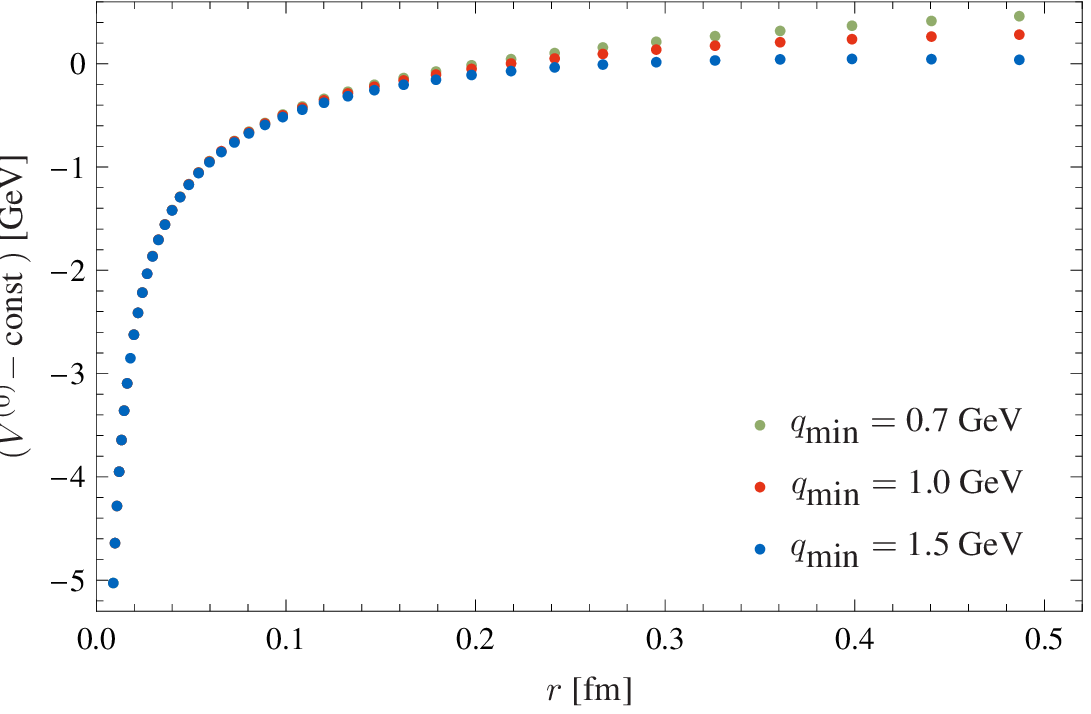}
\hfill
\includegraphics[width=0.48\textwidth]{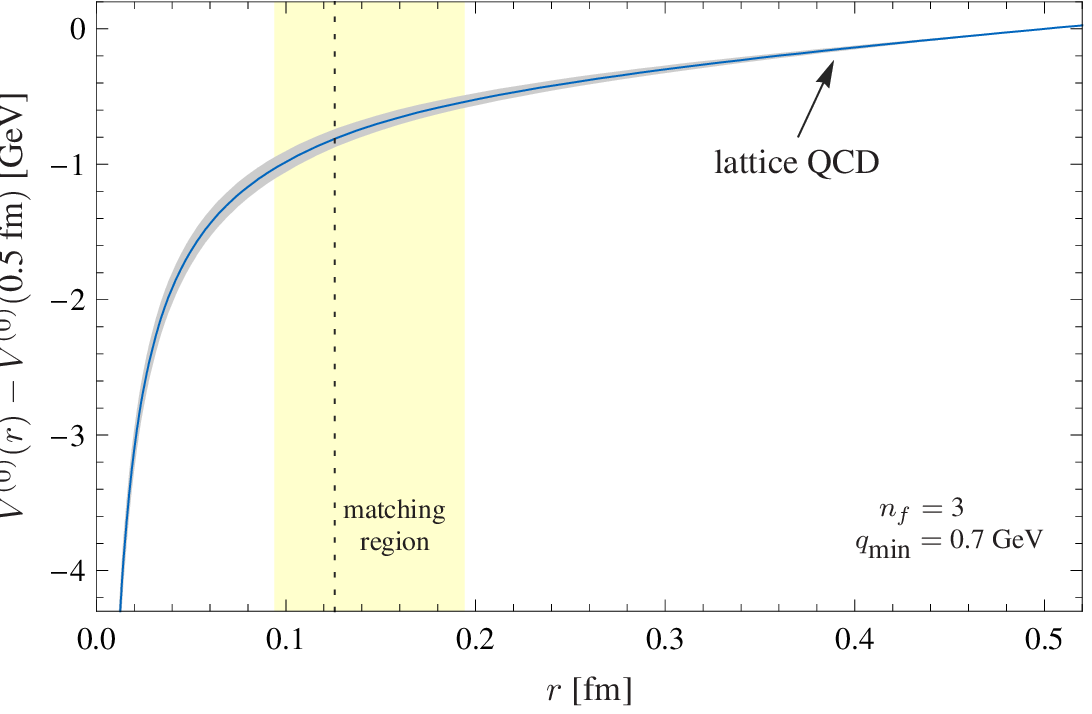}
\end{figure}
In contrast to the usual approach, the running coupling $\alpha_s$ is not
expanded in a power series, but full four-loop RGE
dependence in included in the transformation to coordinate space.
The resulting potential depends only weakly on the cutoff (left figure)
and can be matched at distances $r$ between $0.1$ and $0.2$ fm to a potential
obtained from lattice QCD~\cite{Bali:2000vr}.
The error band of the curve in the right figure reflects uncertainties in the
Sommer scale $r_0= 0.50 \pm 0.03$~fm (lattice part) and uncertainties in the
scale dependence of $\alpha_s(|\vc q|)$ (perturbative part).
The order 1/m potential can be defined analogously and can also be matched
well to calculations from lattice QCD~\cite{Koma:2007jq;Koma:Mainz}.

Work supported in part by BMBF, GSI and by the DFG Excellence Cluster
``Origin and \mbox{Structure} of the Universe''.

\end{document}